\pdfoutput=1

\documentclass[twocolumn,showpacs,amssymb,aps,nofootinbib,floatfix,groupedaddress]{revtex4-1}

\usepackage{epsfig}
\usepackage{hyperref}
\usepackage{comment}
\usepackage{amsmath}
\usepackage{graphicx,color}

\begin{document} 
\hbadness=10000

\title{Principal component analysis of  the nonlinear coupling of harmonic modes 
in heavy-ion collisions}

\author{Piotr Bo\.zek}
\email{piotr.bozek@fis.agh.edu.pl}
\affiliation{AGH University of Science and Technology, Faculty of Physics and
Applied Computer Science, al. Mickiewicza 30, 30-059 Krakow, Poland}

\begin{abstract}
The principal component analysis  of flow correlations in heavy-ion collisions is studied. 
The correlation matrix of harmonic flow is generalized to correlations involving
several different flow vectors. The method can be applied to study the nonlinear coupling between
different harmonic modes in a double differential way in transverse momentum or pseudorapidity. The procedure is illustrated with results from the hydrodynamic model applied to Pb+Pb collisions at $\sqrt{s}=2760$~GeV.
Three examples of  generalized  correlations matrices in transverse momentum  are constructed corresponding
to the coupling of $v_2^2$ and $v_4$, of $v_2v_3$ and $v_5$, or of  $v_2^3$, $v_3^3$, and $v_6$. The principal component decomposition is applied to the correlation matrices and the dominant modes  are calculated.
\end{abstract}

\date{\today}


\keywords{relativistic heavy-ion collisions, collective flow, correlations and fluctuations}

\maketitle


\section{Introduction \label{sec:intro}}

The  expansion of the matter formed  in relativistic heavy-ion collisions generates a collective transverse flow.
 The flow velocity field reflects the gradients in the initial density profile of the fireball.
The harmonic coefficients of the azimuthal asymmetry of the  spectra of emitted particles
 can be measured and compared to model predictions \cite{Heinz:2013th,Gale:2013da,Ollitrault:2010tn}.
The most notable examples are the elliptic $v_2$ and triangular $v_3$ flow coefficients.

Higher flow harmonics $v_n$, $n>3$, are coming from two sources, the expansion of the 
initial asymmetries of the source and due to the nonlinear  coupling of lower order modes  \cite{Teaney:2012ke}.
Correlations between flow harmonics of different order 
have been the subject of numerous theoretical 
\cite{Bhalerao:2011yg,Qiu:2012uy,Jia:2012ma,Jia:2012ju,Teaney:2013dta,Yan:2015jma,Noronha-Hostler:2015dbi,Gardim:2016nrr,Qian:2016fpi,Qian:2017ier,Zhu:2016puf,Giacalone:2016afq,Floerchinger:2013tya,Bravina:2013ora} and experimental studies
\cite{Aad:2014fla,ALICE:2016kpq,Acharya:2017zfg,Adamczyk:2017byf,Tuo:2017ucz}.
One of the motivations was to find additional constraints on the initial state in heavy-ion collisions. 
What is even more important, the sensitivity of the linear and nonlinear response to the viscosity
 of the medium may serve as a tool to estimate of 
the value of shear viscosity in the deconfined quark-gluon plasma.

The correlations of flow harmonics at different transverse momenta \cite{Gardim:2012im} 
or pseudorapidities \cite{Bozek:2010vz} could reveal interesting information on fluctuations 
in the initial state of the evolution. A useful method to analyze the correlation matrix of flow harmonics
is the principal component analysis (PCA) \cite{Bhalerao:2014mua}.
 The procedure separates the leading and subleading 
components in the correlation matrix. The leading component corresponds to the usual flow and the 
subleading component is a measure  of the flow factorization breaking at two different bins in phase space
\cite{Mazeliauskas:2015vea,Mazeliauskas:2015efa,Cirkovic:2016kxt,Sirunyan:2017gyb}.

The correlation matrix for different flow harmonics reflects the mode mixing. Harmonic modes 
at different bins in phase space are partially decorrelated due to factorization breaking. The two effects 
can be 
combined by constructing a more general correlation matrix between different flow harmonics calculated
 at two different bins in phase space. This full correlation matrix can be decomposed into its principal 
components. The procedure analyzes the mixing of different modes in a double differential way in momentum.

I show examples from a hydrodynamic model for the PCA of the correlation matrices corresponding 
to the coupling of $v_2^2$ with $v_4$ and $v_2v_3$ with $v_5$. A correlation matrix of yet higher 
dimension is studied in the example of the correlation between the flow harmonics 
 $v_2^3$, $v_3^2$, and $v_6$.
In semi-central collisions a strong mixing of different modes is found.

\section{Principal component analysis of harmonic flow coefficients \label{sec:pca}}

\begin{figure}[tb]
\includegraphics[width=.48 \textwidth]{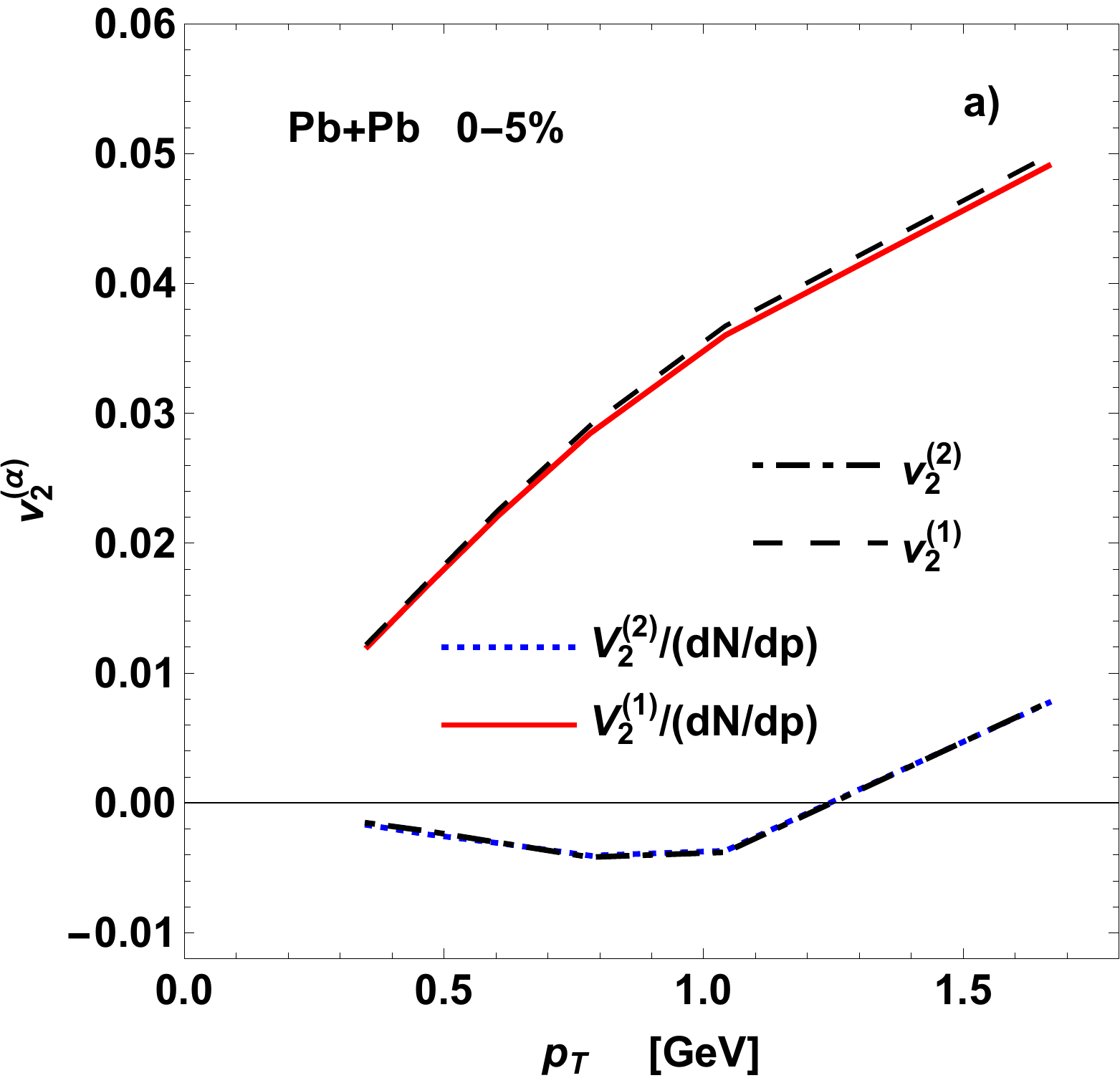}
\vskip 2mm

\includegraphics[width=.48 \textwidth]{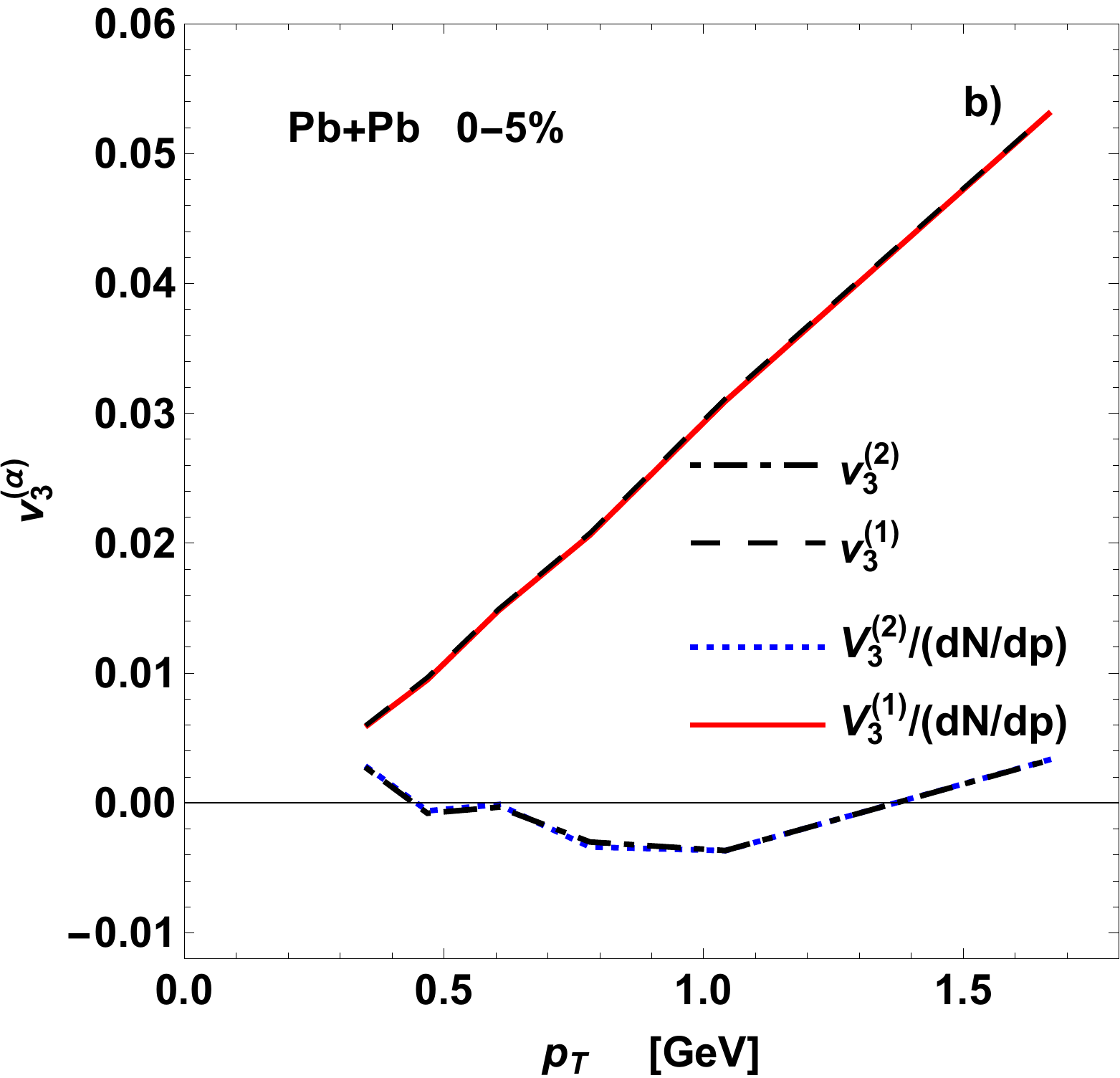}

\caption{First (solid line) and second (dotted line) scaled eigenvectors $V_n^{(\alpha=1,2)}(p)/(dN/dp)$  of the correlation matrix $C_{n\Delta}(p_1,p_2)$ for the second 
(panel a) and third (panel b) order flow in Pb+Pb 
collisions at $\sqrt{s}=2760$~GeV and centrality $0-5$\%. The dashed and dash-dotted lines represents 
the corresponding eigenvectors of the correlation matrix $c_{n\Delta}(p_1,p_2)$. 
\label{fig:vec05}}
\end{figure}

\begin{figure}[tb]
\includegraphics[width=.48 \textwidth]{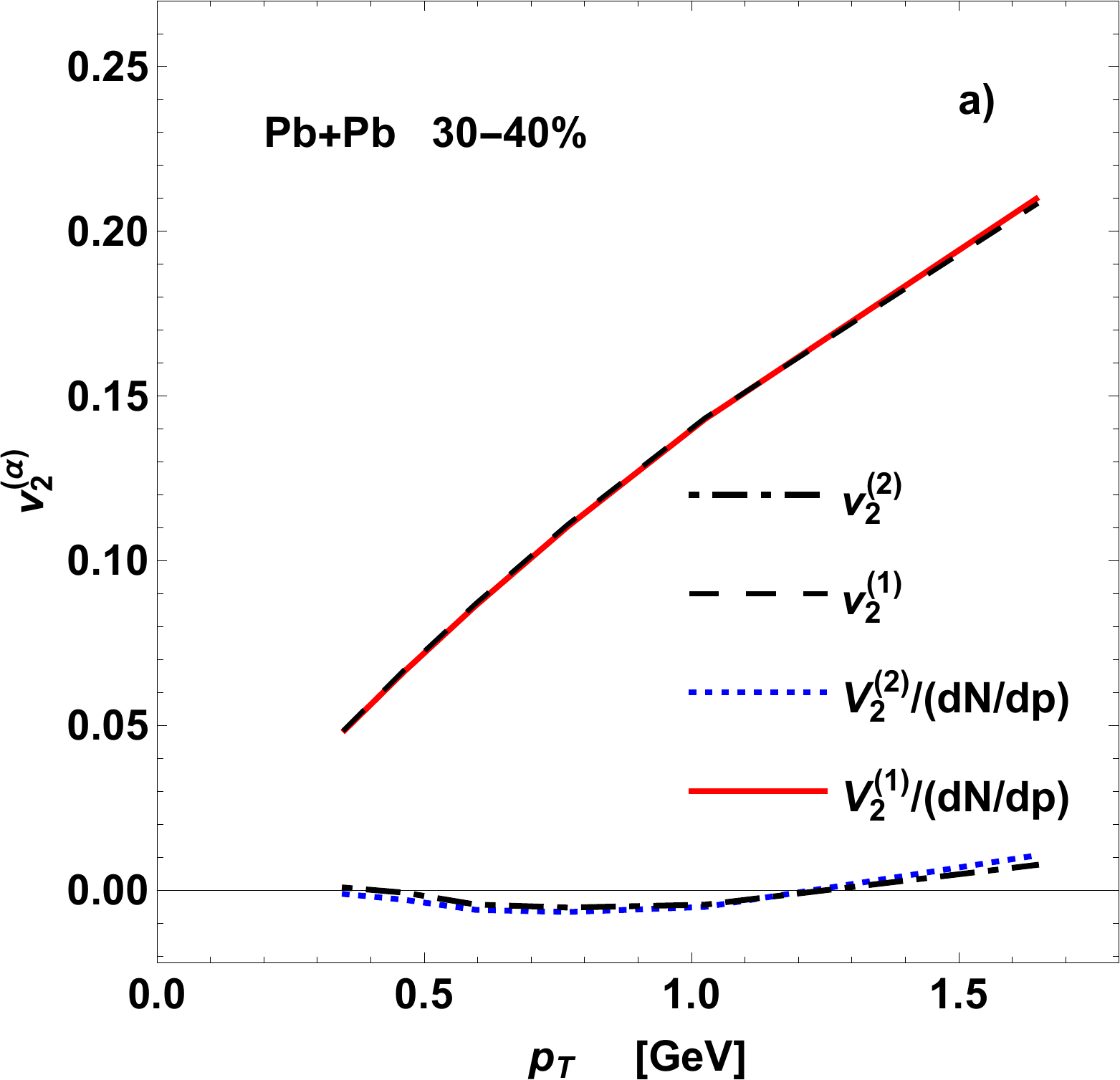}
\vskip 2mm

\includegraphics[width=.48 \textwidth]{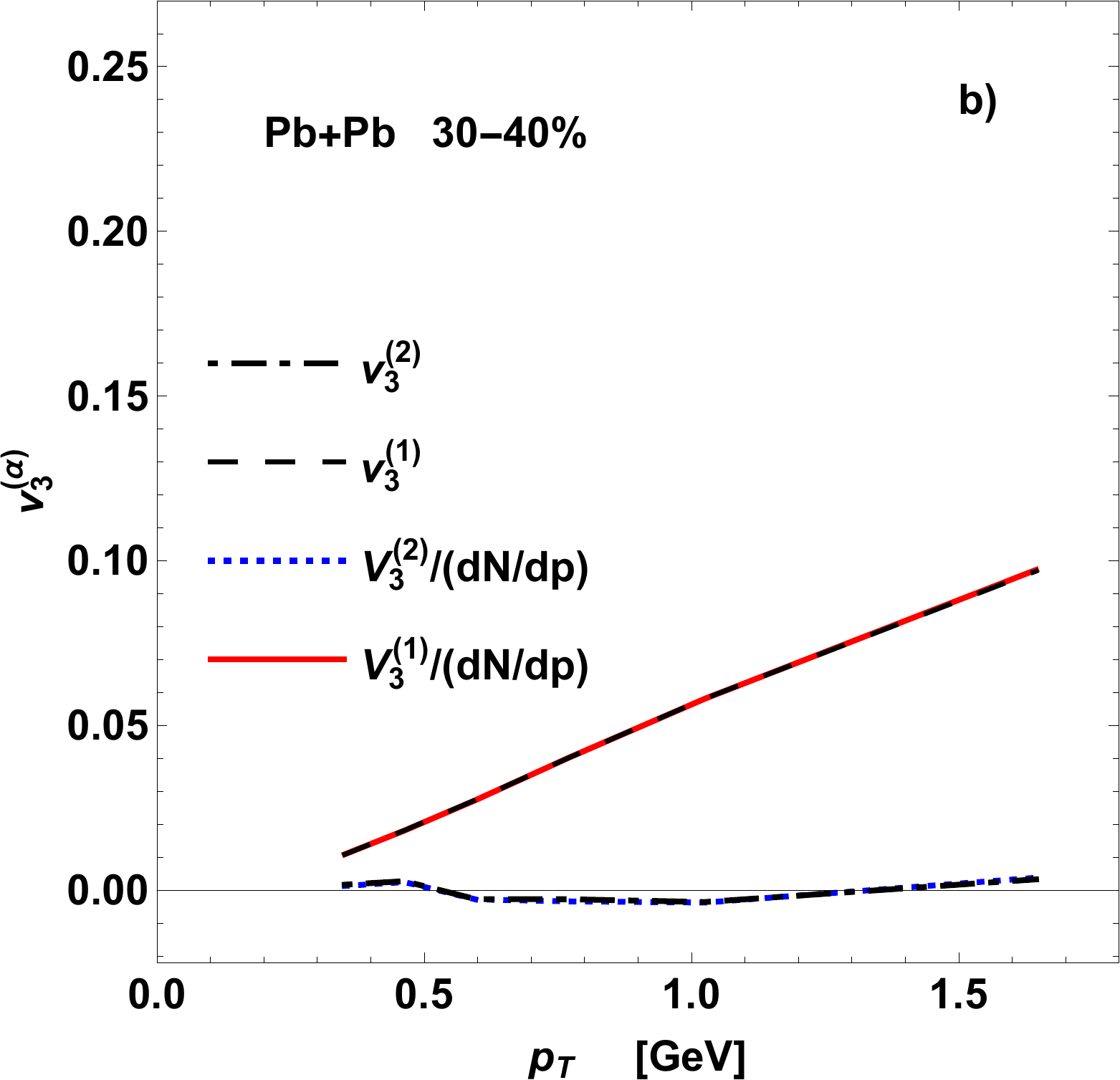}

\caption{Same as in Fig. \ref{fig:vec05} but for Pb+Pb collisions with centrality $30-40$\%.
\label{fig:vec3040}}
\end{figure}

The calculations are performed in a 3+1-dimensional viscous hydrodynamic model with event-by-event
 fluctuating initial conditions \cite{Schenke:2010rr,Bozek:2011ua}. The initial conditions are taken from
a Glauber Monte Carlo model with quark degrees of freedom. The details of the calculation can be found
 in \cite{Bozek:2017elk}. After the hydrodynamic expansion, particles are
 emitted statistically from the freeze-out hypersurface \cite{Chojnacki:2011hb}.

For each hydrodynamic event many events are generated according to statistical emission from the freeze-out hypersurface. This allows to 
perform the calculation in two variants, the first
using realistic events with nonflow correlations from resonance
 decays 
and the second 
 using combined events from the same hydrodynamic evolution. The last method reduces fluctuations 
in observables involving several particles, 
which allows to estimate correlations with up to six flow vectors.

The correlation of flow harmonics \cite{Bhalerao:2014mua}
\begin{equation}
C_{n\Delta}(p_1,p_2)=\langle Q_n(p_1) Q_n^\star(p_2) \rangle -\langle Q_n(p_1)\rangle \langle Q_n^\star(p_2) \rangle
\label{eq:Cndelta}
\end{equation}
 is defined as the correlation of two flow vectors
\begin{equation}
Q_n(p)=\sum_j e^{i n\phi_j} \ ,
\end{equation}
where the sum is over all particles in the bin $p$, and $\langle \dots \rangle$ denotes the 
average over the events. The variable $p$ 
is  the pseudorapidity or transverse momentum. The last term on the right hand side of Eq. 
 \ref{eq:Cndelta}
is needed only for  multiplicity correlations, $n=0$. It is implicitly assumed that  when the  
sums over the particles
run over the same bin  ($p_1=p_2$) self-correlation terms are subtracted. This gives an estimate 
of the correlation matrix of the collective flow at momenta $p_1$ and $p_2$.

Correlations of observables depending on transverse momentum are difficult to construct
from the experimental data due to rapidly falling spectra. 
The correlations can be defined using transformed variables, with a flat spectrum \cite{Bialas:1990dk}. 
The transformed variable $0<X<1$ is the cumulative probability for the distribution in transverse momentum 
$\frac{dN}{dp}$, $p_{min}<p<p_{max}$
\begin{equation}
X(p)=\frac{\int_{p_{min}}^{p} dp^{'} \frac{dN}{dp^{'}}}{\int_{p_{min}}^{p_{max}} dp^{'} \frac{dN}{dp^{'}}} \ .
\end{equation}
By definition the distribution $\frac{dN}{dX}$ is flat. The correlation function 
$C_{n\Delta}(X_1,X_2)$ constructed in bins of the cumulative variable has uniform statistical 
uncertainties in all bins. This property
makes the PCA more stable. Technically, the procedure is equivalent to using $k$ unequal bins in transverse 
momentum, corresponding to $k$ quantiles
 of the distribution $\frac{dN}{dp}$. Unless otherwise stated, I use
6 bins for $0.3$~GeV$<p<3$~GeV. For each bin the average $p$ is used to indicate the corresponding 
values on plots. In particular,  the last bin is approximately $[1.2,3.0]$~GeV with the average value $1.66$~GeV
for the centralities studied.

The eigenvalues $\lambda_{n}^{(\alpha)}$ and eigenvectors $\psi_n^{(\alpha)}(p)$ of the correlation matrix $C_{n\Delta}(p_1,p_2)$
are found
\begin{eqnarray}
C_{n\Delta}(p_1,p_2) & = & \sum_{\alpha} \lambda_n^{(\alpha)} \psi_n^{(\alpha)}(p_1)\psi_n^{(\alpha)}(p_2) \nonumber \\
& = & \sum_\alpha V_n^{(\alpha)}(p_1)V_n^{(\alpha)}(p_2) \ .
\label{eq:cneig}
\end{eqnarray}
Finally, eigenvectors are scaled by the multiplicity distribution 
$ dN/dp$ as in  \cite{Bhalerao:2014mua}. 

For correlations of flow coefficients  $n>0$ the correlation of flow vectors normalized by the multiplicity
\begin{equation}
q_n=\frac{1}{n}\sum_j e^{i n\phi_j}
\end{equation}
 can be constructed
\begin{equation}
c_{n\Delta}(p_1,p_2)=\langle q_n(p_1) q_n^\star(p_2)\rangle  \ .
\label{eq:cndelta}
\end{equation}
As before, for the same bin $(p_1=p_2)$ self-correlations are avoided, and the  normalization 
is by the number of pairs in the bins. The definition of the correlation matrix using the normalized $q_n$
 vectors is useful for comparison to some models where  the spectra and the 
collective flow is calculated without generating realistic  finite multiplicity  events.
In the following the correlation matrices with upper case $C$  and lower case $c$ denote correlation 
of flow vectors $Q$ and $q$ respectively.
The eigenvectors of correlation matrix $c$ are denoted with a lower case letter  $v_n^{(\alpha)}(p)$
\begin{equation}
c_{n\Delta}(p_1,p_2) =\sum_\alpha v_{n}^{(\alpha)}(p_1) v_{n}^{(\alpha)}(p_2) \ .
\end{equation}
Note that there is no need to scale the eigenvectors $v_n^{(\alpha)}(p)$  by the average multiplicity.

The scaled eigenvectors of the harmonic flow correlations of second and third order are shown in Figs.
\ref{fig:vec05} and \ref{fig:vec3040} for Pb+Pb collisions with centralities $0-5$\% and $30-40$\% 
respectively.
The largest eigenvalue is dominant in the decomposition of the correlations matrix. It is consistent
 with the small factorization breaking of flow coefficient in transverse momentum \cite{Bhalerao:2014mua}.
On the same figures are plotted the eigenvectors of the correlation matrices $c_{n\Delta}(p_1,p_2)$
for the multiplicity normalized $q_n(p)$ vectors. The results are very similar to the eigenvectors of the
correlations matrices $C_{n\Delta}(p_1,p_2)$.

\section{Principal component analysis of coupled flow harmonics}

Harmonics of different order 
can be coupled due to correlations in the initial distribution of eccentricities 
or due to nonlinearities in the evolution 
\cite{Bhalerao:2011yg,Yan:2015jma,Noronha-Hostler:2015dbi,Gardim:2016nrr,Qian:2016fpi,Qian:2017ier,Zhu:2016puf,Giacalone:2016afq}. The mixed flow harmonics are usually calculated in a given  
acceptance region.
More generally any mixed flow harmonic can be estimated using  different bins in momentum.
The simplest class of such correlators \cite{Bhalerao:2011yg} defined for two bins in momentum is
\begin{widetext}
\begin{eqnarray}
v\{n_1 \dots n_l ,- m_1 \dots -m_k\} (p_1,p_2)& = & \langle Q_{n_1}(p_1) \dots Q_{n_l}(p_1) Q_{m_1}^\star(p_2) \dots Q_{m_k}^\star(p_2) \rangle \nonumber \\ & = &
\langle \sum_{j_1 \neq \dots \neq j_l} \sum_{ s_1 \neq \dots \neq  s_k}  e^{i(n_1 \phi_{j_1} +\dots + n_l 
\phi_{j_l} - m_1 \phi_{s_1} - \dots -
 m_k \phi_{s_k} )}  \rangle
\label{eq:vnn}
\end{eqnarray}
the first and second sums 
 run over particles in bins $p_1$ and $p_2$ respectively and $N=\sum_{i=1}^l n_i = \sum_{j=1}^k m_k$ .
The above formula defines a (in general  asymmetric) correlation matrix in momenta. Mode mixing 
suggests to study  correlations of  different possible harmonic of order  $N$  in the  bin $p_1$ and of order  $N$ 
 in the bin $p_2$. The full correlation matrix of order involves all such possible combinations
of flow harmonics
\begin{eqnarray}
C_{n\dots | m\dots| q\dots | \dots }(p_1^{I},p_1^{II},p_1^{III},\dots|p_1^{I},p_1^{II},p_1^{III},\dots) & = & \nonumber \\
\begin{bmatrix}
v\{n\dots,-n\dots\}(p_1^{I},p_2^{I}) & v\{n\dots,-m\dots\}(p_1^{I},p_2^{II}) & v\{n\dots,-q\dots\}(p_1^{I},p_2^{III}) & \dots \\
v\{m\dots,-n\dots\}(p_1^{II},p_2^{I}) & v\{m\dots,-m\dots\}(p_1^{II},p_2^{II}) & v\{m\dots,-q\dots\}(p_1^{II},p_2^{III}) & \dots \\
v\{q\dots,-n\dots\}(p_1^{III},p_2^{I}) & v\{q\dots,-m\dots\}(p_1^{IIII},p_2^{II}) & v\{q\dots,-q\dots\}(p_1^{IIII},p_2^{III}) & \dots \\
\dots & \dots & \dots & \dots
\end{bmatrix} \ .
\label{eq:cgeneral}
\end{eqnarray}
The matrix $C_{n\dots | m\dots| q\dots | \dots }(p_1^{I},p_1^{II},p_1^{III},\dots|p_1^{I},p_1^{II},p_1^{III},\dots)$
is symmetric. Note that the dimension of the correlation matrix is the 
multiple of the dimension of the simple
$C_{n\Delta}$ correlation matrices; it is indicated by the multiple momentum indices for the rows and columns
 of the matrix. For flow dominated correlations 
$C_{n\dots | m\dots| q\dots | \dots }(p_1^{I},p_1^{II},p_1^{III},\dots|p_1^{I},p_1^{II},p_1^{III},\dots)$
is positive semi-definite.

The most interesting correlations involve harmonic modes with strong coupling. In particular,
from the most general form of the correlation matrix of order $N$ a submatrix can be chosen for
 which the off-diagonal terms in Eq. \ref{eq:cgeneral} are significant.
In the following are listed few examples of such correlation matrices.

The coupling of $v_2^2$ harmonic to $v_4$ is defined by the matrix 
\begin{equation}
C_{2;2|4}(p_1^{I},p_1^{II}|p_{2}^{I},p_2^{II})=
  \begin{bmatrix}
 \langle Q_2(p_1^{I})^2 Q_2^\star(p_2^{I})^2  \rangle &  \langle Q_2(p_1^{I})^2 Q_4^\star(p_2^{II}) \rangle \\ 
 \langle Q_4(p_1^{II}) Q_2^\star(p_2^{I})^2 \rangle &  \langle Q_4(p_1^{II})Q_4^\star(p_2^{II}) \rangle  
  \end{bmatrix}
\end{equation}
As discussed in section \ref{sec:pca} the analogous correlation for normalized $q$ vector is
\begin{equation}
c_{2;2|4}(p_1^{I},p_1^{II}|p_{2}^{I},p_2^{II})=
  \begin{bmatrix}
 \langle q_2(p_1^{I})^2 q_2^\star(p_2^{I})^2  \rangle &  \langle q_2(p_1^{I})^2 q_4^\star(p_2^{II}) \rangle \\ 
 \langle q_4(p_1^{II}) q_2^\star(p_2^{I})^2 \rangle &  \langle q_4(p_1^{II})q_4^\star(p_2^{II}) \rangle  
  \end{bmatrix}
\label{eq:c224} \  .
\end{equation}
The nonlinear coupling  of $v_2v_3$ and $v_5$ shows up in the correlation matrix
\begin{equation}
c_{23|5}(p_1^{I},p_1^{II}|p_{2}^{I},p_2^{II})= 
  \begin{bmatrix}
 \langle q_2(p_1^{I}) q_3(p_1^{I}) q_2^\star(p_2^{I}) q_3^\star(p_2^{I})  \rangle &  \langle q_2(p_1^{I}) q_3(p_1^{I}) q_5^\star(p_2^{II}) \rangle \\ 
 \langle q_5(p_1^{II}) q_2^\star(p_2^{I}) q_3^\star(p_2^{I})  \rangle &  \langle q_5(p_1^{II})q_5^\star(p_2^{II}) \rangle  
  \end{bmatrix}
\label{eq:c235}\ .
\end{equation}
A more complicated matrix with coupling in three sectors $v_2^3$, $v_3^2$, and $v_6$ is
\begin{equation}
c_{2;3|3;2|6}(p_1^{I},p_1^{II},p_1^{III}|p_{2}^{I},p_2^{II},p_2^{III})=  
  \begin{bmatrix}
 \langle q_2(p_1^{I})^3  q_2^\star(p_2^{I})^3   \rangle &  \langle q_2(p_1^{I})^3 q_3^\star(p_2^{II})^2\rangle 
& \langle q_2(p_1^{I})^3 q_6^\star(p_2^{III}) \rangle \\ 
 \langle q_3(p_1^{II})^2  q_2^\star(p_2^{I})^3   \rangle &  \langle q_3(p_1^{II})^2 q_3^\star(p_2^{II})^2\rangle 
& \langle q_3(p_1^{II})^2 q_6^\star(p_2^{III}) \rangle \\ 
  \langle q_6(p_1^{III})  q_2^\star(p_2^{I})^3   \rangle &  \langle q_6(p_1^{III}) q_3^\star(p_2^{II})^2\rangle 
& \langle q_6(p_1^{III}) q_6^\star(p_2^{III}) \rangle 
  \end{bmatrix} \ .
\label{eq:c23326}
\end{equation}
Any submatrix  of a more general matrix   operators correlations  can be  considered
\begin{equation}
c_{2;3|6}(p_1^{I},p_1^{II}|p_{2}^{I},p_2^{II})= 
  \begin{bmatrix}
 \langle q_2(p_1^{I})^3  q_2^\star(p_2^{I})^3   \rangle &  
 \langle q_2(p_1^{I})^3 q_6^\star(p_2^{II}) \rangle \\ 
  \langle q_6(p_1^{II})  q_2^\star(p_2^{I})^3   \rangle
& \langle q_6(p_1^{II}) q_6^\star(p_2^{II}) \rangle 
   \end{bmatrix}
\label{eq:c236}
\end{equation}
or
\begin{equation}
c_{3;2|6}(p_1^{I},p_1^{II}|p_{2}^{I},p_2^{II})= 
  \begin{bmatrix}
\langle q_3(p_1^{I})^2 q_3^\star(p_2^{I})^2\rangle 
& \langle q_3(p_1^{I})^2 q_6^\star(p_2^{II}) \rangle \\ 
  \langle q_6(p_1^{II}) q_3^\star(p_2^{I})^2\rangle 
& \langle q_6(p_1^{II}) q_6^\star(p_2^{II}) \rangle 
  \end{bmatrix} \ .
\label{eq:c326}
\end{equation}

The correlation matrices can be decomposed in eigenvectors in the same way as the correlation $c_{n\Delta}(p_1,p_2)$
\begin{eqnarray}
c_{\dots}(p_1^{I},p_1^{I},\dots|p_{2}^{I},p_2^{II},\dots)& = & \sum_{\alpha} \lambda_{\dots}^{(\alpha)} \psi_{\dots}^{(\alpha})(p_1^{I},p_1^{II},\dots)\psi_{\dots}^{(\alpha)}(p_2^{I},p_2^{II},\dots) \nonumber \\
& = & \sum_{\alpha} v_{\dots}^{(\alpha)}(p_1^{I},p_1^{II},\dots)v_{\dots}^{(\alpha)}(p_2^{I},p_2^{II},\dots) \ ,
\label{eq:decfull}
\end{eqnarray}
\end{widetext}
with
$v_{\dots}^\alpha(p_1^{I},p_1^{II},\dots)=\sqrt{\lambda_{\dots}^\alpha} \psi_{\dots}^\alpha(p_1^{I},p_1^{II},\dots)$.
Note that the eigenvectors have a higher  dimension than for the correlations $c_{n\Delta}(p_1,p_2)$, e.g.
\begin{equation}
v^{(\alpha)}_{2;2|4}(p^{I},p^{II})=
\begin{bmatrix}
v^{(\alpha,I)}_{2;2|4}(p^{I}) \\
v^{(\alpha,II)}_{2;2|4}(p^{II})
\end{bmatrix}
\end{equation}
or
\begin{equation}
v^{(\alpha)}_{2;3|3;2|6}(p^{I},p^{II},p^{III})=
\begin{bmatrix}
v^{(\alpha,I)}_{2;3|3;2||6}(p^{I}) \\
v^{(\alpha,II)}_{2;3|3;2||6}(p^{II}) \\
v^{(\alpha,III)}_{2;3|3;2||6}(p^{III})
\end{bmatrix}  \ .
\end{equation}

In the limit, when the mixing of
 different harmonic components in $c_{\dots}$ is negligible the dominant components of the first 
eigenvectors $v_{\dots}^{(q)}$ are close to the moments the relevant harmonic operators.
For example the eigenvectors of $c_{2;2|4}$ would be
\begin{equation}
v_{2;2|4}^{(\alpha)}(p^{I},p^{II}) \simeq \begin{bmatrix}
0  \\
\sqrt{\langle v_4(p^{II})^2 \rangle}
\end{bmatrix}
\end{equation}
 or  
\begin{equation}v_{2;2|4}^{(\alpha)}(p^{I},p^{II}) \simeq \begin{bmatrix}
\sqrt{\langle v_2(p^{I})^4 \rangle}\\
0 
\end{bmatrix} \ , 
\end{equation}
for $\alpha=1,2$.

\section{Hydrodynamic model results \label{sec:cgen}}


\begin{figure}
\includegraphics[width=.48 \textwidth]{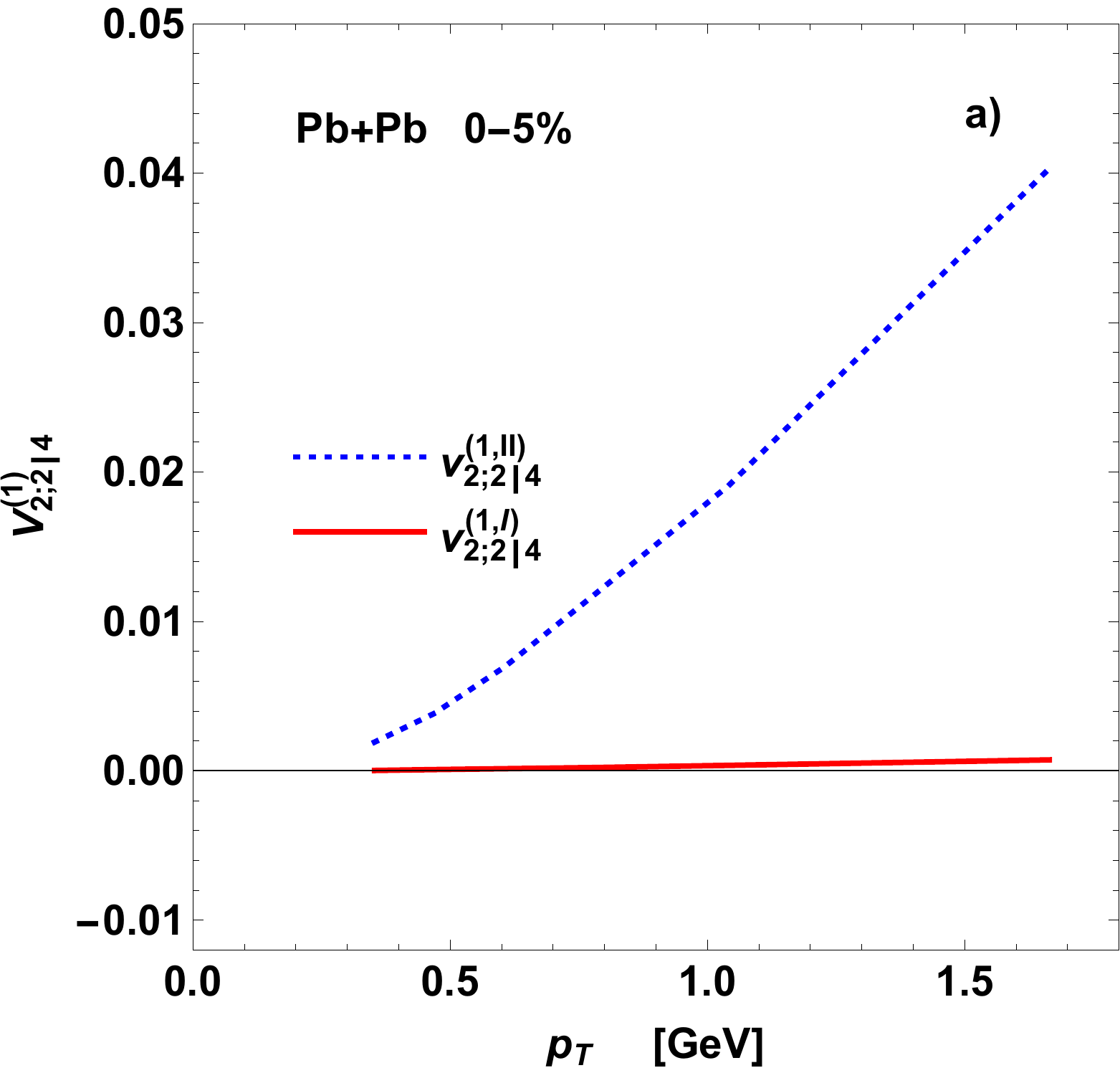}
\vskip 2mm

\includegraphics[width=.48 \textwidth]{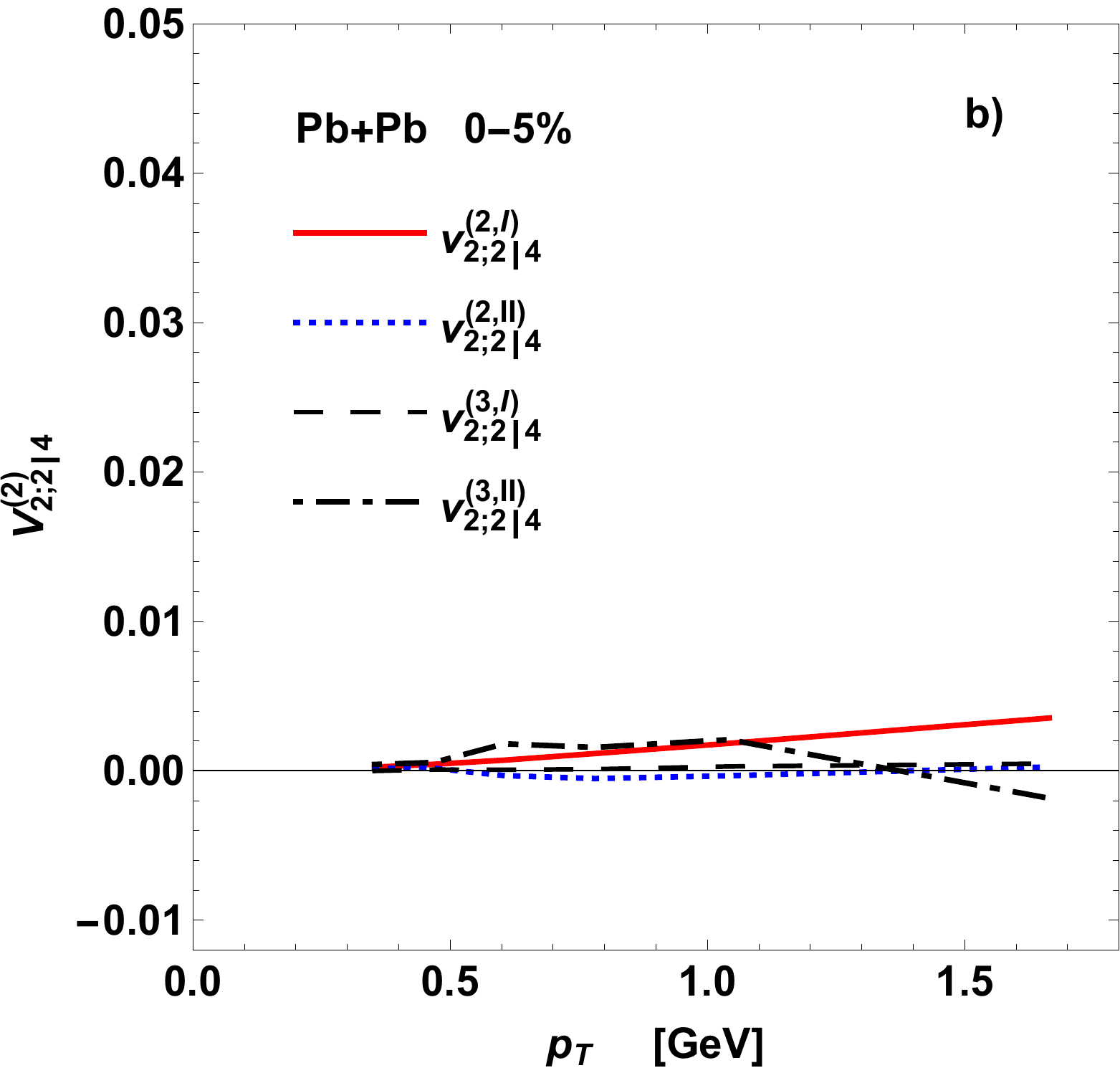}

\caption{Eigenvectors of the correlation matrix $c_{2;2|4}$ (Eq. \ref{eq:c224}). Panel a), the two components of the first eigenvector; panel b), the two components of the second and third eigenvectors; Pb+Pb collisions with 
centrality $0-5$\%.
\label{fig:vec22i405}}
\end{figure}

\begin{figure}
\includegraphics[width=.48 \textwidth]{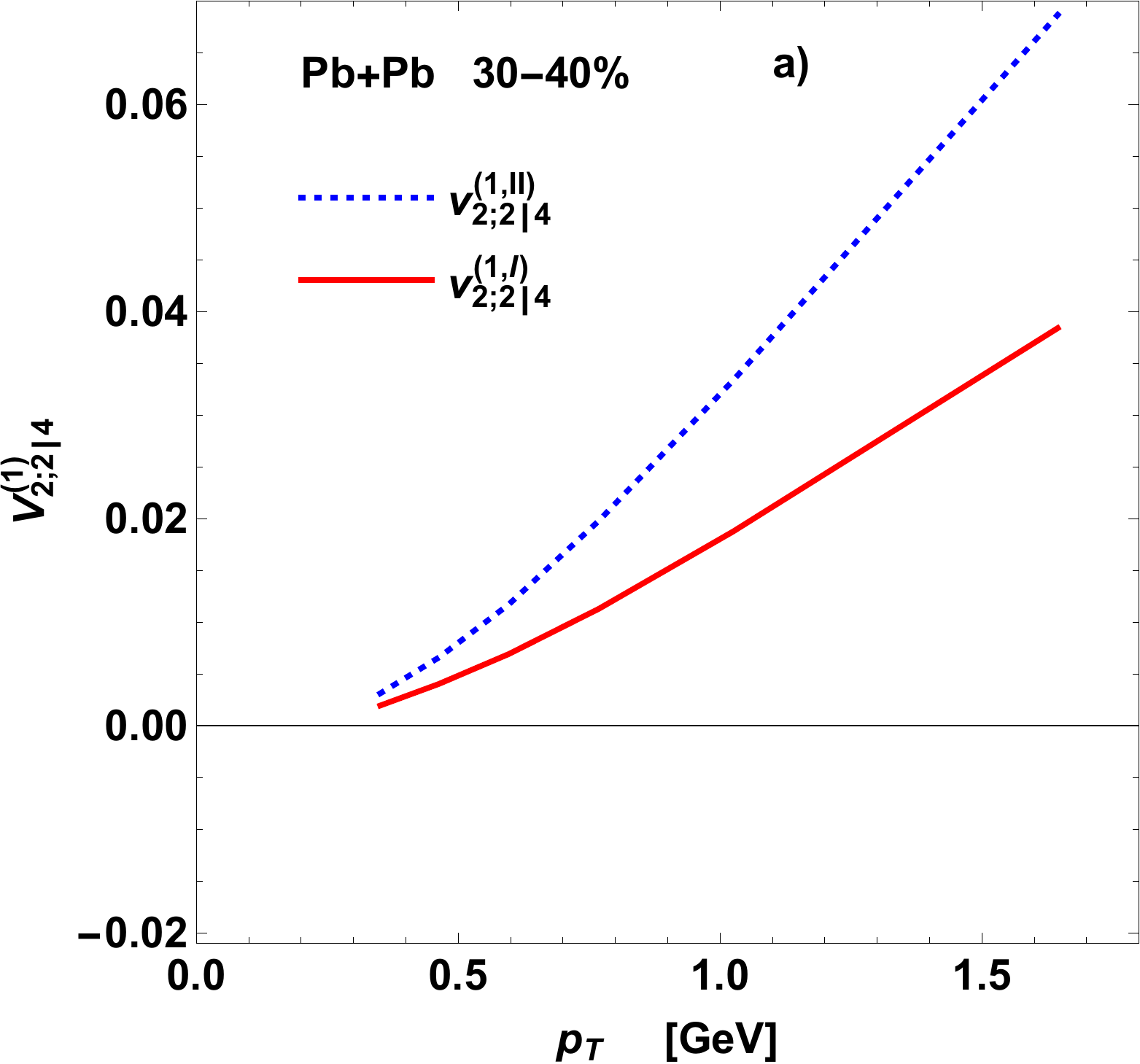}
\vskip 2mm

\includegraphics[width=.48 \textwidth]{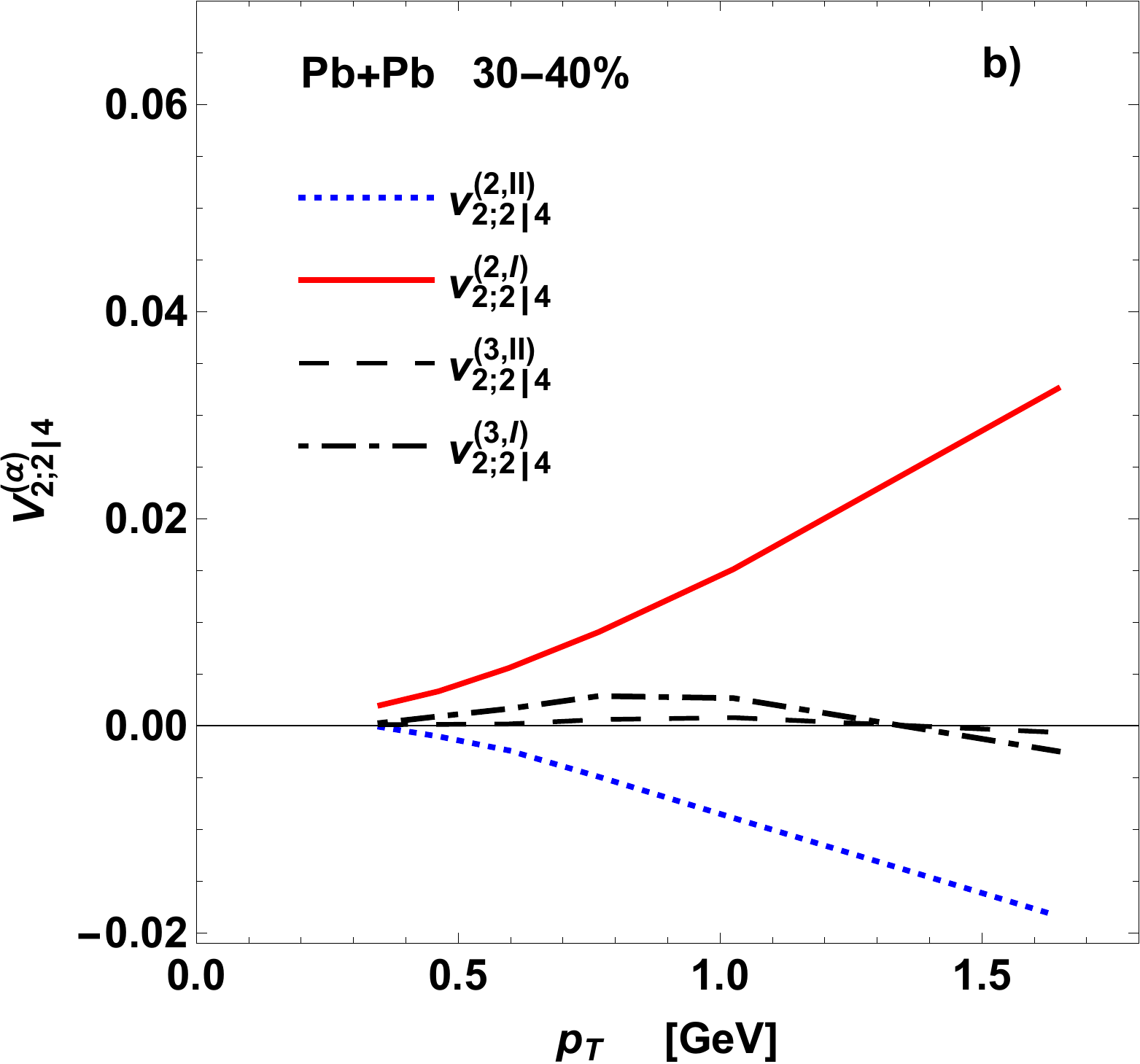}

\caption{Same as in Fig. \ref{fig:vec22i405} but for collisions with centrality $30-40$\%.
\label{fig:vec22i43040}}
\end{figure}

The hydrodynamic results presented in this section serve as an illustration of the possibility to
perform the PCA for the generalized correlation matrices. Extensive, high statistics simulations for
 different initial conditions and viscosities are beyond the scope of this paper. The correlations
 in this section are calculated combining many  events in same the hydrodynamic evolution.  Thus, 
statistical fluctuations and nonflow effects are reduced.

The nonlinear coupling between $v_2^2$ and $v_4$ is expected to be significant whenever 
the elliptic flow is strong. The results of the decomposition in principal components for the
correlator $c_{2;2|4}$ (Eq. \ref{eq:c224}) are shown in Figs. \ref{fig:vec22i405} and \ref{fig:vec22i43040}.
 For central collisions $0-5$\% the nonlinear effects are small. The leading mode is located
 in the sector $v_6$  (Fig. \ref{fig:vec22i405} panel a)). 
The subleading mode is located in the sector $v_2^2$ with   small mixing to $v_6$  
(Fig. \ref{fig:vec22i405} panel b)). Only 
in the third mode a momentum dependent factorization breaking effect shows up clearly in the sector $v_6$.

The mixing of the modes   $v_2^2$ and $v_4$ is stronger for semi-central collisions $30-40$\%.
The leading and the subleading modes have significant components
 in both sectors (Fig. \ref{fig:vec22i43040}). This mixing can be understood as due to a nonlinear 
contribution to the $v_4$ flow \cite{Teaney:2012ke}
\begin{equation}v_4 = v_4^L + \chi_{422} v_2^2 \ .
\end{equation}
The PCA of the correlation matrix $c_{22;2|4}$  in momentum takes into account 
this nonlinear coupling, while being sensitive to possible
additional effects of factorization breaking $\langle v_2(p_1) v_2(p_2)\rangle \neq
 \sqrt{\langle v_2(p_1)^2 \rangle \langle  v_2(p_2)^2\rangle}$ and of a momentum 
dependence of the coupling $\chi_{422}$. The third  mode $v_{2;2|4}^{(3)}$ shows a momentum dependent 
factorization breaking, mainly in the sector $v_4$.


\begin{figure}
\includegraphics[width=.48 \textwidth]{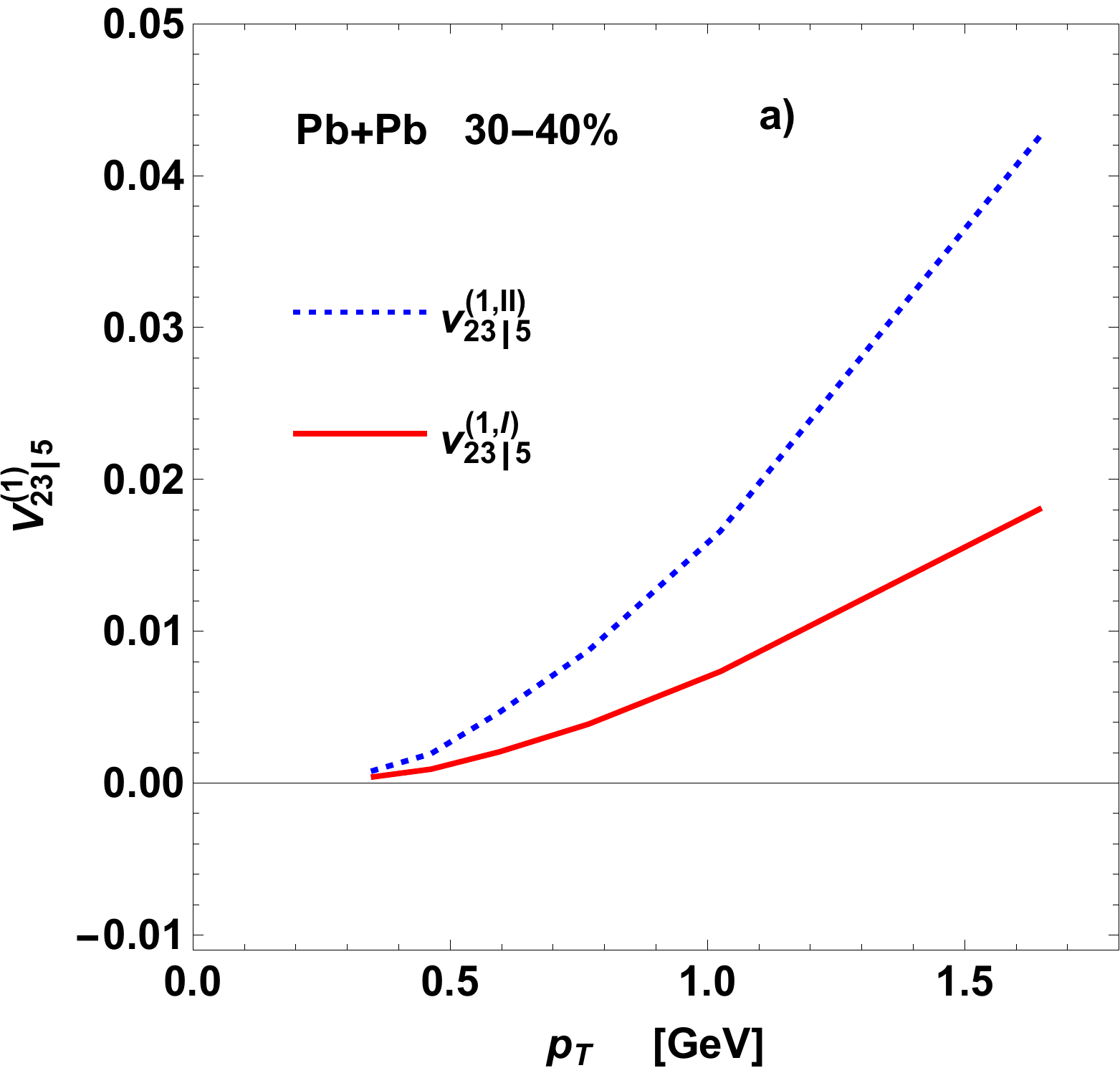}
\vskip 2mm

\includegraphics[width=.48 \textwidth]{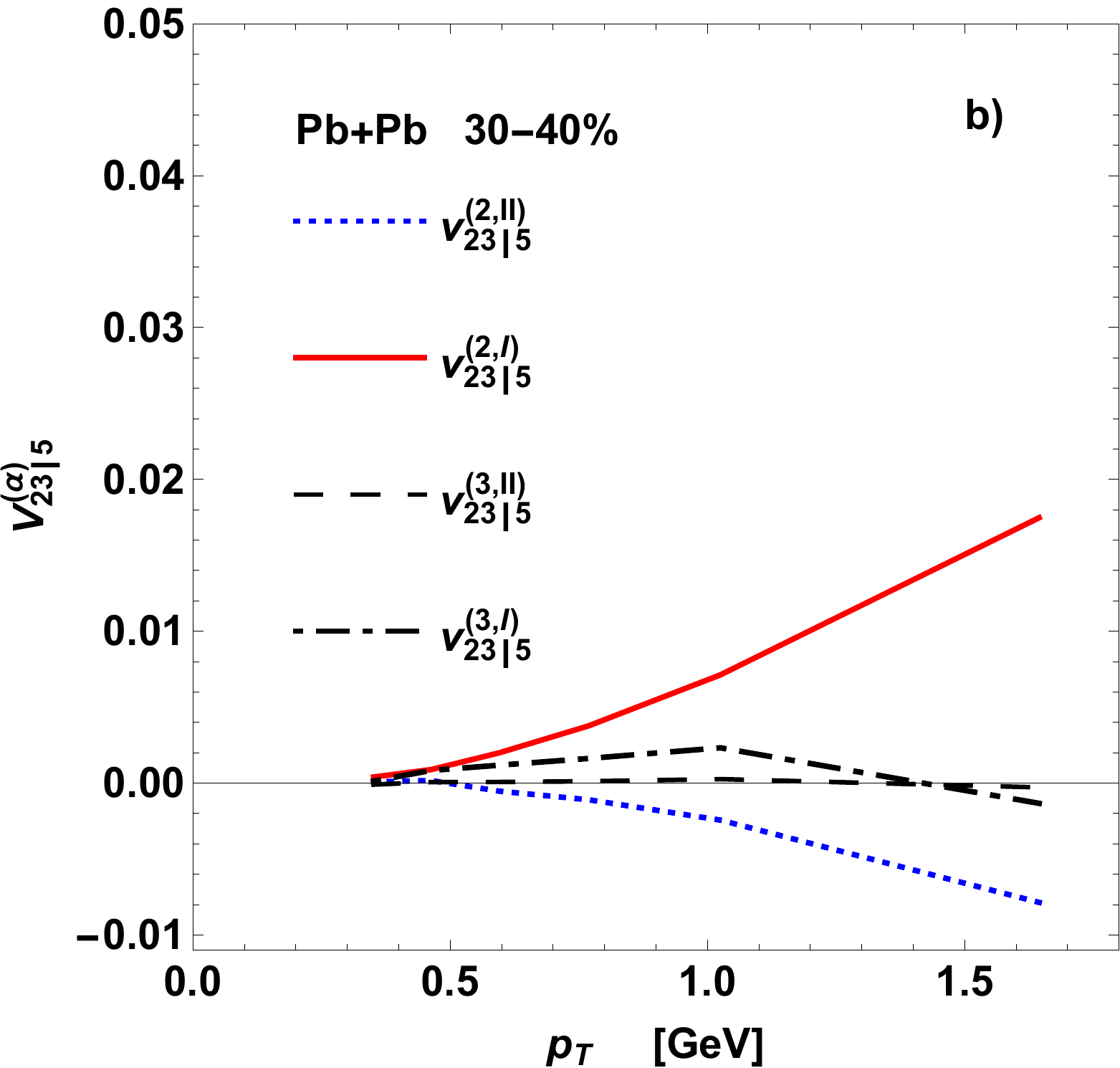}

\caption{Same as in Fig. \ref{fig:vec22i43040} for the correlation matrix $c_{23|5}$.
\label{fig:vec23i53040}}
\end{figure}

The flow $v_5$ get a contribution trough nonlinear coupling from the harmonic flow $v_2v_3$.
In Fig. \ref{fig:vec23i53040} are shown the first three eigenvectors for the correlation matrix
$c_{23|5}$ (Eq. \ref{eq:c235}). The results are qualitatively similar to the case of  $v_2^2$-$v_4$ coupling.
The mixing between the modes $v_2v_3$ and $v_5$ is strong  for the first two eigenvectors.
The third eigenvector reflects the factorization breaking in momentum.



\begin{figure*}
\includegraphics[width=.48 \textwidth]{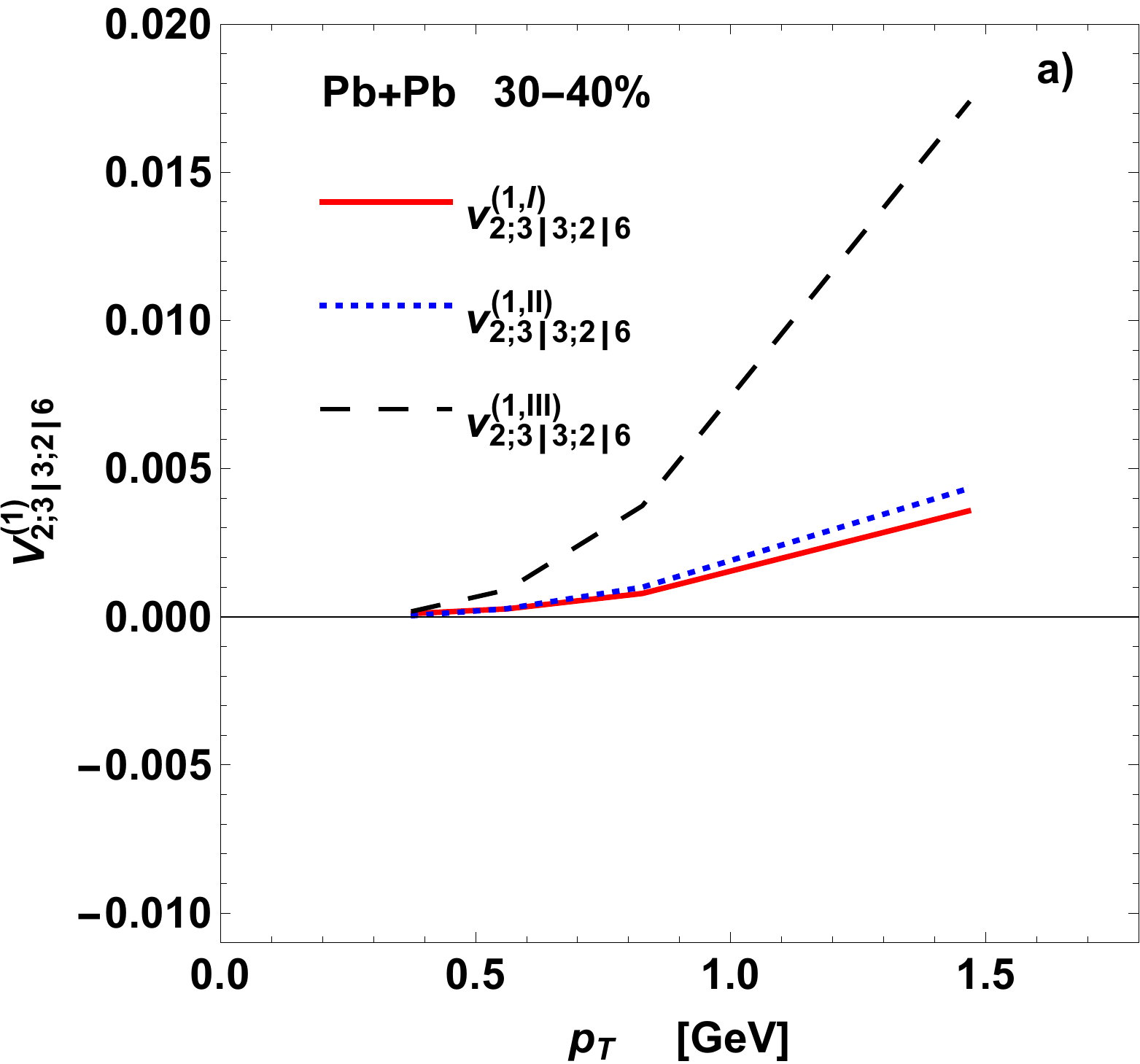}~~~~~\includegraphics[width=.48 \textwidth]{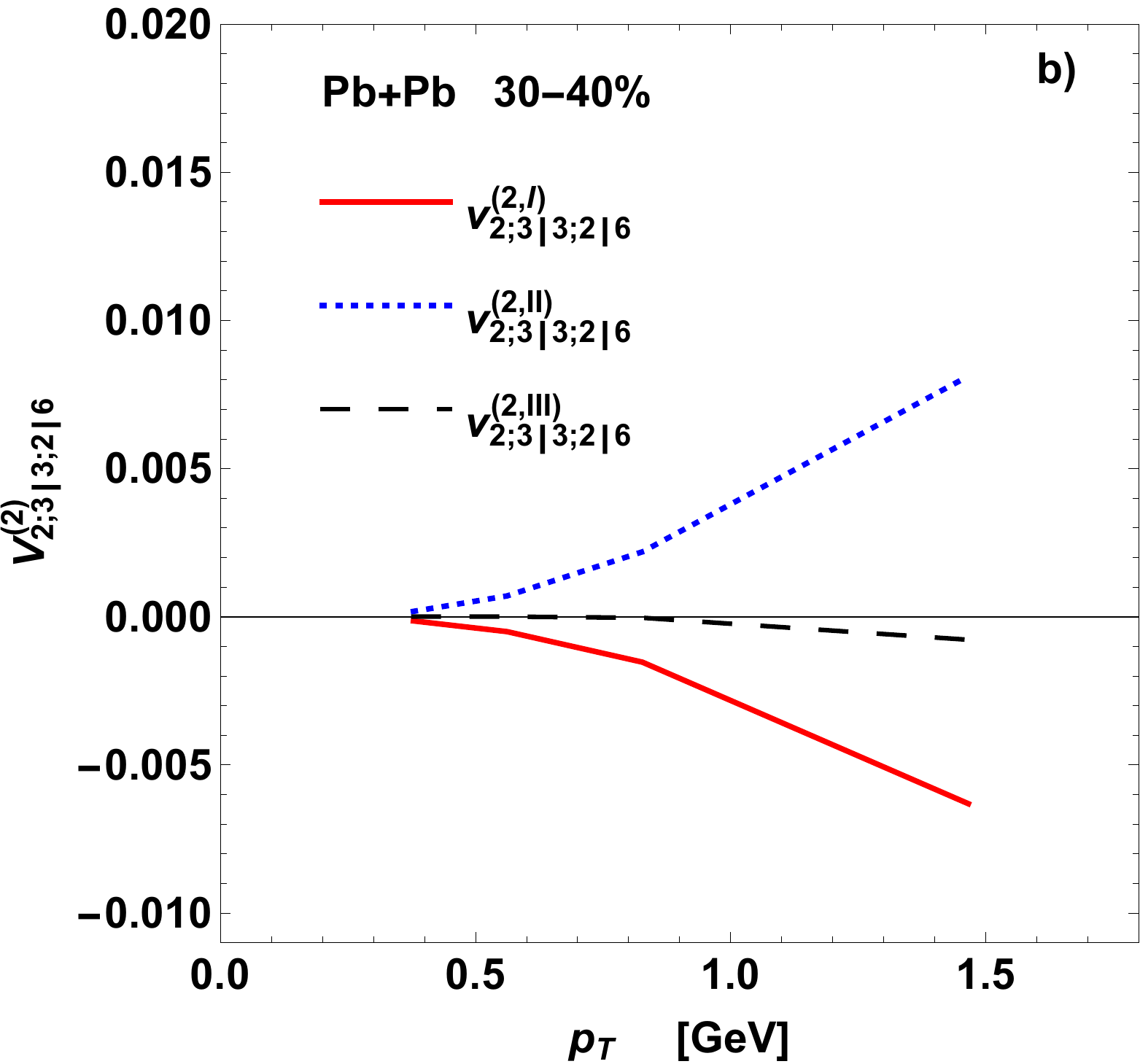}
\vskip 2mm

\includegraphics[width=.48 \textwidth]{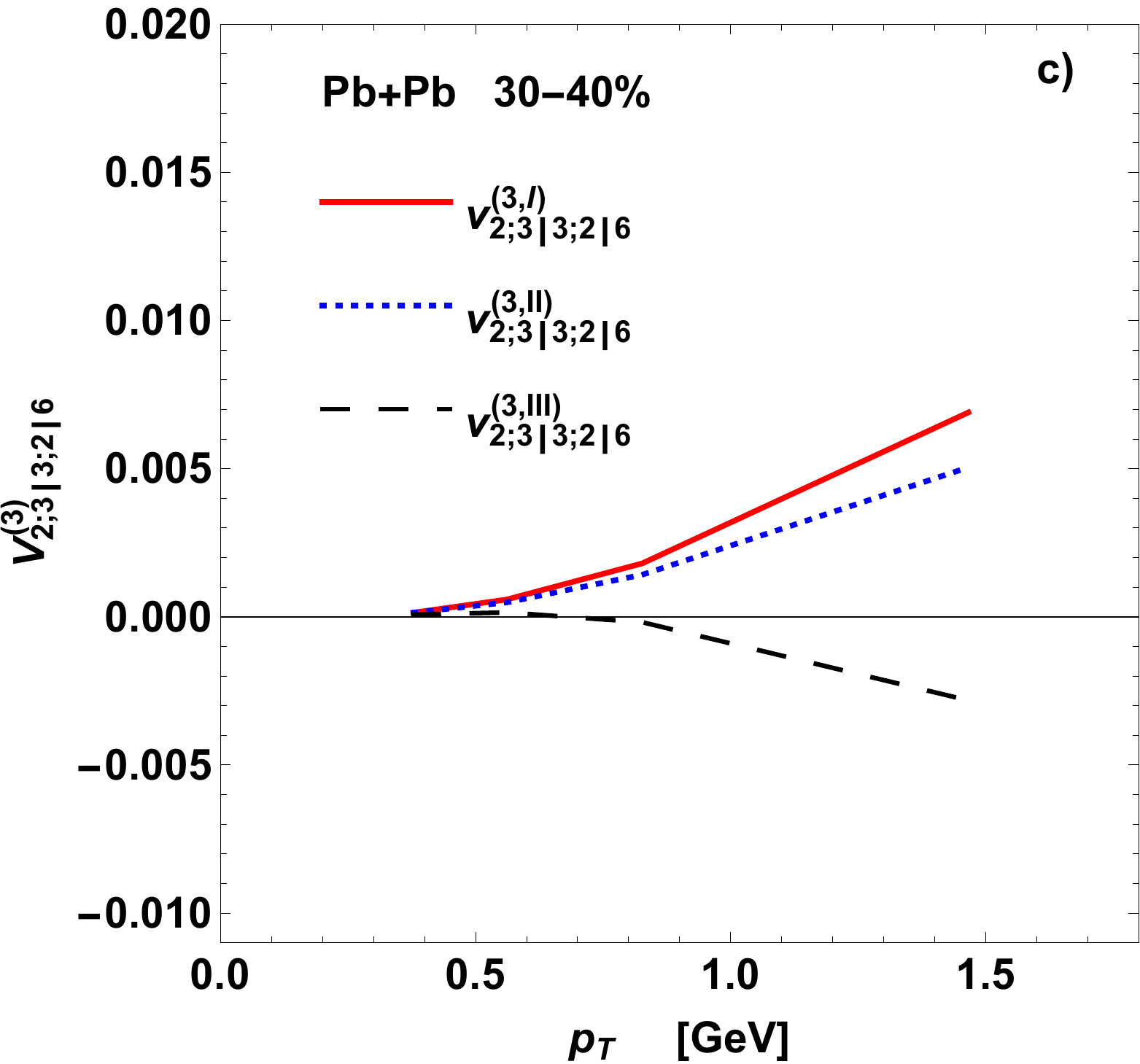}~~~~~\includegraphics[width=.48 \textwidth]{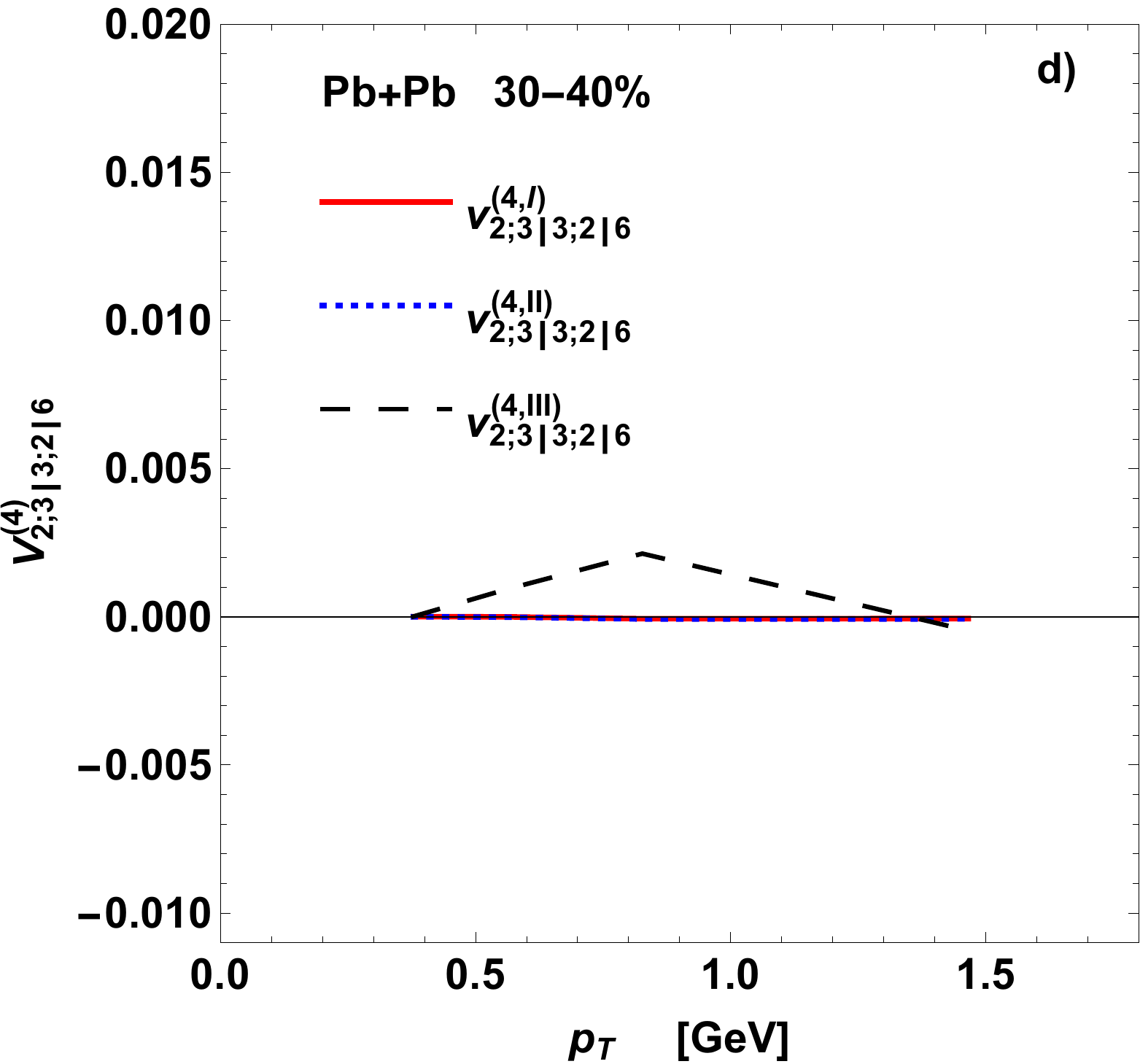}

\caption{Eigenvectors of the correlation matrix $c_{2;3|3;2|6}$. In panels a) through d) are shown the components of the eigenvectors corresponding to the first four  eigenvalues. In all panels three components of the eigenvectors are shown I (solid lines), II (dotted lines), and III (dashed lines) corresponding to the sectors $q_2(p)^2$, $q_3(p)^3$, and $q_6(p)$ in 
the correlation matrix. 
\label{fig:vec23i63040}}
\end{figure*}


The correlation matrix $c_{2;3|3;2|6}$ combines the flow vectors from three sectors $v_2^3$, $v_3^2$, and 
$v_6$. This correlation is calculated using 4 bins in transverse momentum only. As before the unequal bins 
correspond to 4 quantiles of the distribution $dN/dp$ in the range $[0.3,\ 3]$~GeV. The correlation matrix
$c_{2;3|3:2|6}$ has dimensions $12\times 12$, with three sectors $v_2^3$, $v_3^2$, and $v_6$.

The  first four  eigenvectors are plotted in panels a) through d) in Fig. \ref{fig:vec23i63040}.
The leading mode is a mixing in all three sectors with the strongest component from $v_6$. The second and third
 modes show a strong mixing in the $v_2^3$ and $v_3^2$ sectors with almost equal strength. The fourth mode
 shows a momentum dependent response \cite{Mazeliauskas:2015vea,Mazeliauskas:2015efa}, mainly in the $v_6$ sector.

The coupling between different harmonic modes is usually studied for the momentum integrated
flow \cite{Giacalone:2016afq,Yan:2015jma,Qian:2016fpi,Zhu:2016puf}, 
with the exception of Ref. \cite{Qian:2017ier}. In the present formalism, it corresponds
to correlation matrices constructed with only one bin in 
transverse momentum. For the examples studied one has $2\times 2$ or $3\times 3$ matrices
\begin{equation}
c_{2;2|4}=
\begin{bmatrix}
\langle q_2^2 q_2^{\star 2} \rangle & \langle q_2^2 q_4^{\star} \rangle  \\
\langle q_4 q_2^{\star 2} \rangle & \langle q_4 q_4^{\star} \rangle 
\end{bmatrix}
\label{eq:onec224} \ , 
\end{equation}
\begin{equation}
c_{23|5}=
\begin{bmatrix}
\langle q_2 q_3 q_2^{\star } q_3^\star \rangle & \langle q_2 q_3  q_5^{\star} \rangle  \\
\langle q_5 q_2^{\star }q_3^\star \rangle & \langle q_5 q_5^{\star} \rangle 
\end{bmatrix}
\label{eq:onec235} \ , 
\end{equation}
and
\begin{equation}
c_{2;3|3;2|6}=
\begin{bmatrix}
\langle q_2^3 q_2^{\star  3 }  \rangle & \langle q_2^3  q_3^{\star  2} \rangle &  \langle q_2^3  q_6^{\star } \rangle\\
\langle q_3^2 q_2^{\star 3 }  \rangle & \langle q_3^2  q_3^{\star 2} \rangle &  \langle q_3^2  q_6^{\star } \rangle\\
\langle q_6 q_2^{\star 3 } \rangle & \langle q_6  q_3^{\star  2 } \rangle &  \langle q_6  q_6^{\star } \rangle\\
\end{bmatrix}
\label{eq:onec236} \ .
\end{equation}

The components  of the  
two eigenvectors of the matrices $c_{2;2|4}$, $c_{23|5}$ agree qualitatively with
 the hierarchy of the components of the first two eigenvectors of the momentum dependent correlation 
matrices shown 
in Figs. \ref{fig:vec22i405}, \ref{fig:vec22i43040}, and \ref{fig:vec23i53040}. Analogously, the hierarchy 
 of the component of the three eigenvectors of $c_{2;3|3;2|6}$ (Eq. \ref{eq:onec236})
 reflects qualitatively the hierarchy of the components I, II, and III
of the eigenvectors $v_{2;3|3;2|6}^{(\alpha)}$ in Fig. \ref{fig:vec23i63040} for $\alpha=1,2,3$.

\section{Conclusions}

The flow pattern in heavy-ion collisions reflects on event-by-event basis the fluctuations and correlations
present in the initial state as well as the development of different modes in the hydrodynamic
 evolution. The study of flow harmonics
 and their correlations gives constraints both on the initial state and  on the
properties of the expanding dense matter in the fireball. I propose to study the 
correlations between different flow harmonics at two different transverse momenta
 (or pseudorapidities). 

The full correlation matrix has a block structure. Each block can be either a correlation matrix 
between the same flow harmonics at two different momenta or a correlations of two different
 flow harmonics at two different momenta. The PCA is performed on this generalized 
correlation matrix. One finds
 momentum dependent eigenmodes corresponding to the modes resulting 
from the mixing of different harmonics. For a correlation matrix build out of $n$ different harmonics
 the first $n$ eigenvectors reflect mostly the mode mixing. Only higher eigenvectors show the 
 weak component due to factorization breaking.

Model simulations and experimental measurements of the correlation
 matrix between different flow harmonics yield additional information on  harmonic 
mode mixing in heavy-ion collisions. The PCA performed in transverse momentum is a method to study
 mode mixing at different momenta and factorization breaking in the same framework. 
A double differential study in transverse momentum or rapidity could also identify  
sources of mode mixing other than collective flow.

\vskip 7mm

\begin{acknowledgments}

Research supported by the Polish Ministry of Science and Higher Education (MNiSW), by the National
Science Centre grant  2015/17/B/ST2/00101, as well as by PL-Grid Infrastructure. 

\end{acknowledgments}

\bibliography{../hydr}

\end{document}